\documentclass[]{spie}
\usepackage[dvips]{color}
\usepackage{aastex_hack}
\usepackage{amssymb}
\usepackage{epsfig}
\usepackage[latin1]{inputenc}
\usepackage{graphicx}
\usepackage{times}
\usepackage{amssymb}
\usepackage{color}
\usepackage[figuresright]{rotating}
\usepackage{amsmath}
\usepackage{url}

\newcommand{\figurewidth} {0.40\textwidth}

\title{Faint flux performance of an EMCCD}
\author{Olivier Daigle\supit{a,b}, Claude Carignan\supit{a}, Sébastien Blais-Ouellette\supit{b}
\skiplinehalf
\supit{a} Laboratoire d'Astrophysique Exp\'erimentale, D\'epartement de physique, Universit\'e de Montr\'eal, C.P. 6128 Succ. Centre-Ville, Montr\'eal, QC, Canada, H3C 3J7;\\
\supit{b} Photon etc., 5155 Decelles Avenue, Pavillon J.A Bombardier, Montr\'eal, Qu\'ebec, Canada, H3T 2B1.}

\authorinfo{Send correspondance to O.D.: odaigle@astro.umontreal.ca}

\begin{document}
\maketitle

\begin{abstract}
Thorough numerical simulations were run to test the performance of three processing methods of the data coming out from an electron multiplying charge coupled device (EMCCD), or low light level charge coupled device (L3CCD), operated at high gain, under real operating conditions. The effect of read-out noise and spurious charges is tested under various low flux conditions (0.001 event/pixel/frame $< f <$ 20 events/pixel/frame). Moreover, a method for finding the value of the gain applied by the EMCCD amplification register is also developed. It allows one to determine the gain value to an accuracy of a fraction of a percent from dark frames alone.\\
\end{abstract}
\keywords{Astronomical instrumentation, EMCCD, L3CCD, IPCS}


\section{INTRODUCTION}
\label{sect:intro}
The advent of electron multiplying charge coupled devices (EMCCD) allows one to apply a gain to the pixel's charge before it reaches the noisy output amplifier\cite{techreport-minimal2}. Sub-electron read-out noise levels are thus reachable. However, this kind of signal amplification comes to a price: the stochastic multiplication process induces a noise on the gain level that renders impossible to determine the exact gain that has been applied to a pixel's charge. This statistical behaviour thus adds a noise factor that reaches a value of $2^{1/2}$ at high gains\cite{stanford}. The effect on the signal-to-noise ratio (SNR) of the system is the same as if the quantum efficiency (QE) of the CCD would be halved.

Some signal processing techniques may allow one to overcome this noise and recover the full silicon QE of the CCD. Simulations are run to test the behaviour of different processing techniques. The simulations and their results are presented in section \ref{sect:simulations}.

Throughout the processing of the signal coming out of an EMCCD, the mean gain of the EM register is a key value that has to be determined with a high accuracy if one wants to acquire absolute photometric information with such a device. In section \ref{sect:data}, an algorithm is developed so that the real gain of the EMCCD can be calculated through the processing of dark frames alone.

\section{SIMULATIONS}
\label{sect:simulations}
Thorough numerical simulations are run to properly understand the effect of the multiplication register on the pixel signal. This enables the optimization of the signal processing in order to recover the original pixel's value. Simulations concentrated mostly on low fluxes, 20 photons/pixel/frame being the highest flux simulated.

The simulation process is divided into four sections:
\begin{enumerate}
\item The generation of the pixel's signal. This includes the photon's signal as well as the spurious charges generated during the vertical transfer;
\item The journey of the pixel's charge into the EMCDD multiplication register;
\item The simulation of the output amplifier's noise that is added to the multiplied pixel's charge;
\item The digital processing of the output signal to try to recover the amount of input photons.
\end{enumerate}

\subsection{Generating the pixel's signal}
The generation of the pixel's signal involves generating a Poissonian flux of a given intensity. Then, if one wants to simulate the behaviour of the spurious charges, a Poissonian intensity of spurious charges is generated for every pixel. The simulations cover only the frame interval and not the time interval. Thus, these simulations do not take into account the dark signal, which have a time component. As for the spurious charges, only the clock induced charges (CIC) are simulated, which have a frame component.

\subsection{Simulating the multiplication register}
The simulation of the EMCCD multiplication gain is done as follows. If one wants to simulate a multiplication register of $n$ elements having a mean gain of $g = \overline{G}$, a probability of multiplication of $p = g^{1/n}-1$ must be used for every element of the register. Then, the multiplication register is expanded as follows:

\begin{enumerate}
\item The electrons contained into a pixel are loaded into the first stage of the multiplication register;
\item Every electron is given the probability \textit{p} of being multiplied;
\item For every electron that is multiplied, another electron is added to the pool of electrons;
\item The resulting electrons are loaded into the next stage of the multiplication register and the algorithm loops to step 2 until all the elements of the multiplication were passed through;
\item At the end, the amount of electrons corresponds to the output value of the multiplication register. 
\end{enumerate}

\begin{figure*}
\begin{center}
\includegraphics[width=\figurewidth]{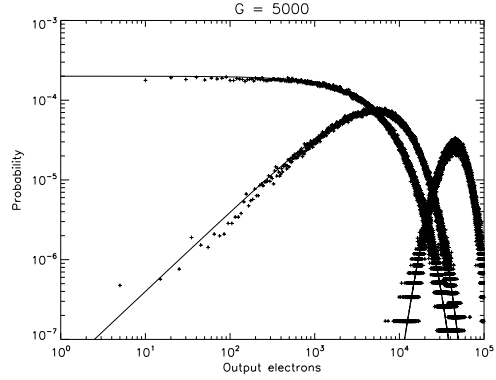}
\caption{Simulated (crosses) and theoretical (plain lines) output values of an EMCCD multiplication register of 536 elements for a mean gain of 5000. The plotted values are for one, two and ten input electrons (leftmost to rightmost curve). The simulated values were binned into 10 output electrons bins. The hatched appearance of the simulation results at low probability comes from the amount of iterations of the simulations. The generation of more events would be required to overcome this.}
\label{fig:sim_comparison}
\end{center}
\end{figure*}

The algorithm have been cross-checked with the theoretical output value of the multiplication register of an EMCCD, which is
\begin{equation*}
p(x,n,g) = \dfrac{x^{n-1}e^{-x/g}}{g^n(n-1)!},
\end{equation*}
where $x$ and $n$ are respectively the output and input values and $g$ is the mean gain. The results are shown in figure \ref{fig:sim_comparison}. One could argue that there is no need of generating simulated output values for simulating the real behaviour of an EMCCD. However, this approximation is not valid for large $n$, as explained in [\citenum{2003MNRAS.345..985B}]. Thus, one would prefer to do the full simulations in order not to be affected by the approximation.

\subsection{Generating the read-out noise}
The simulation of the output amplifier's noise is just a matter of adding a noise with a normal distribution, that has a standard deviation value of the actual read-out noise, to the output value of the multiplication register. This value is considered as the one being read by the analog to digital converter. This value is then handed over to the digital processing routines.

\subsection{Processing the digitalized signal}
Various output processing strategies are tested. First, for low fluxes, the simulations focus on the performance of the single threshold method (photon counting: PC), where the pixels are considered as being binary: the output is interpreted as being the result of one input photon or none. This method is extensively described in [\citenum{2004SPIE.5499..219D}]. Throughout the simulations, a cut-off level of 5.5$\sigma$ above the read-out noise is used.

Then, for higher fluxes, the behaviour of the Poisson probability (PP) thresholding scheme described in [\citenum{2003MNRAS.345..985B}] is tested. This processing strategy was developed to minimize the noise factor of the EMCCD induced by the multiplication register. The simulations allow one to test this affirmation. This multi-threshold scheme uses different threshold boundaries steps. These values are recopied in table \ref{table:pp} for convenience (taken from table 1 of [\citenum{2003MNRAS.345..985B}]).

\begin{table}[htbp]
\begin{center}
\begin{tabular}{cc|cc}
\hline
Threshold & Boundary & Threshold & Boundary\\
\hline
1 & 0.71 & 7 & 6.97\\
2 & 1.89 & 8 & 7.98\\
3 & 2.93 & 9 & 8.98\\
4 & 3.95 & 10 & 9.98\\
5 & 4.96 & 11 & 11.0\\
6 & 5.97 & $n\ge11$ & $n$\\
\end{tabular}
\caption{Threshold boundaries in units of mean gain, for the PP thresholding strategy\cite{2003MNRAS.345..985B}.}
\label{table:pp}
\end{center}
\end{table}%

Finally, a scheme where the amount of photons is simply taken as being the output value divided by the mean gain is tested. This is often referred to as the \textit{analogic} mode.

\subsection{Optimization}
The simulation of the multiplication register being the most CPU demanding task (especially at high gain), a pool of output values of the multiplication register were pre-generated for many input pixel's values, for different mean gains. That is, a million output values were generated for an input of 1 electron and 10000 output values for an input of 100 electrons, with every input values in between. These output values were then stored into a file. Later, when the output value of the multiplication register is needed for a given input value, (between 0 and 100 electrons), it is just a matter of randomly picking a value that was already computed. This proved to substantially speedup the simulations. The amount of pre-generated output values was high enough not to affect the simulation results. No more than half of the pre-generated output values were used for every simulation.

\section{Simulation results}
The simulations were done using different parameters for the read-out noise and the spurious charges. The simulations are run with these values:
\begin{enumerate}
\item Read-out noise 0 electron, spurious charges 0 electron/pixel/frame. This first simulation allows one to see the effect of the electron multiplication process and the processing alone. It is a good indicator as which of the processing methods are best suited for a given flux;
\item Read-out noise 50 electrons, spurious charges 0 electron/pixel/frame. This second simulation allows one to check how the thresholding of the read-out noise performs;
\item Read-out noise 50 electrons, spurious charges 0.06 electron/pixel/frame. This third simulation is run to simulate the real behaviour of an EMCCD, knowing that the spurious charges will be an important source of noise.
\end{enumerate}

In order to analyse the simulations, the effective noise factor (ENF) induced by the different digital processings is considered. The ENF is simply defined as
\begin{equation*}
ENF = \dfrac{\sigma_{out}}{\sigma_{in}},
\end{equation*}
where $\sigma_{out}$ and $\sigma_{in}$ are the standard deviations of the output signal and input signal, respectively. The input signal is always considered as being the photon flux, so it always has a standard deviation of $\sqrt{f}$, where $f$ is the mean flux in photon/pixel/frame.

Next, the \textit{misfit}, the mean error that a processing method induces, is computed. It is defined as
\begin{equation*}
M = \dfrac{\sum_i{\left( n_i - y_i \right)^2}}{\sum_i{n_i}},
\end{equation*}
where $n_i$ is the signal input (Poissonian photons) and $y_i$ is the photon count as determined by the processing method.

Finally, the behaviour of the SNR for the different processing methods is studied. The SNR of the simulations is defined as being
\begin{equation*}
SNR = \dfrac{\left<n\right>}{\sqrt{\left<n\right> + \left<\left(n - y\right)^2\right>}}.
\end{equation*}
In this case, the value of $\left<n\right>$ represents the shot noise and the value of $\left<\left(n - y\right)^2\right>$ represents all the other sources of noise, including the spurious charges and read-out noise. For the thresholding (PC, PP) strategies, this SNR equation is valid only when the gain has a value at least 10 times the read-out noise. The SNR is compared to a perfect photon counting system, where the SNR is expressed as
\begin{equation*}
SNR = \dfrac{\left<n\right>}{\sqrt{\left<n\right>}}.
\end{equation*}

As these simulations aimed at characterizing the different processing strategies between them, they do not take into account the quantum efficiency of the CCD.

\subsection{Effects of the multiplication process}
\label{sect:em_gain}
\begin{figure*}
\begin{center}
\includegraphics[width=\figurewidth]{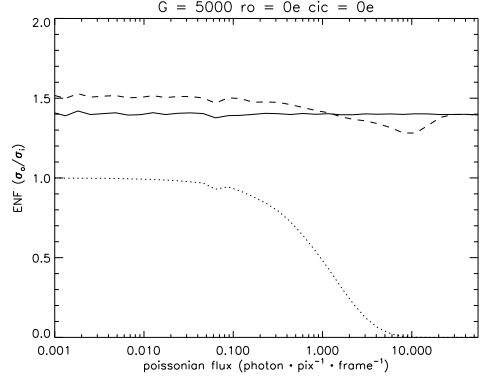}
\caption{Effective noise factor induced by the multiplication process, using different processing methods. Dotted line: PC, dashed line: PP, plain line: analogic.}
\label{fig:enf}
\end{center}
\end{figure*}

\begin{figure*}
\begin{center}
\includegraphics[width=\figurewidth]{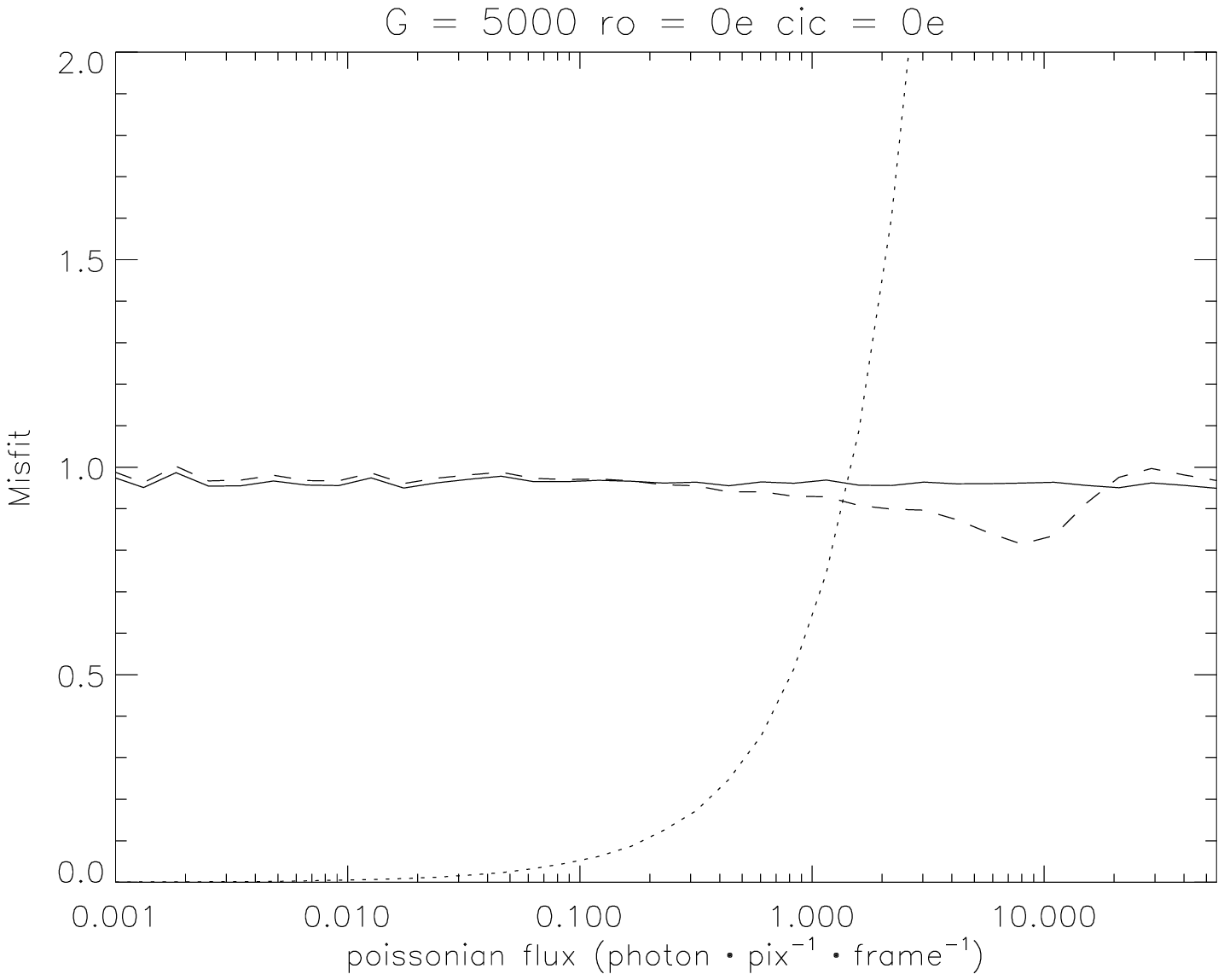}
\includegraphics[width=\figurewidth]{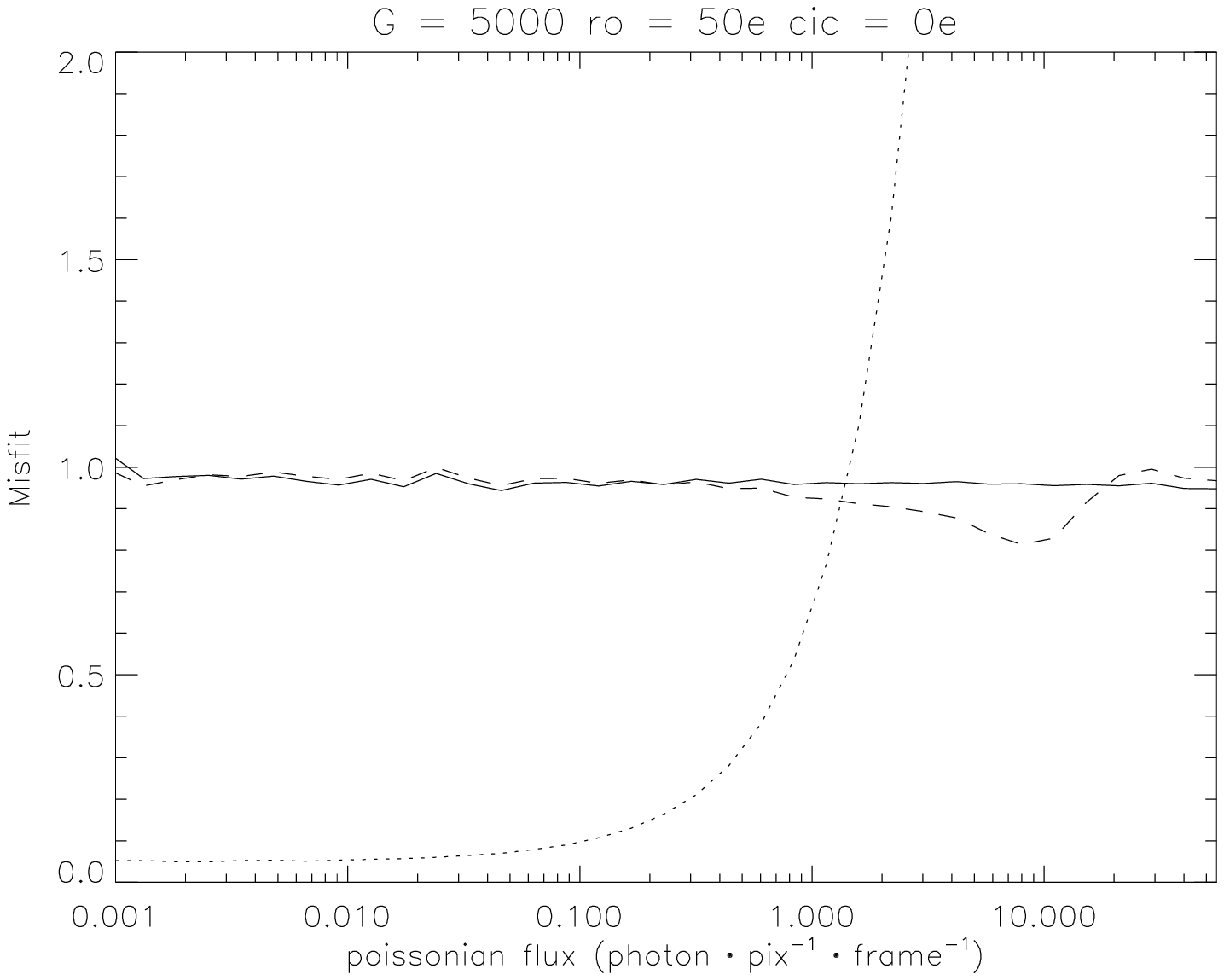}
\caption{Effect of the read-out noise on the misfit . Dotted line: PC, dashed line: PP, plain line: analogic.}
\label{fig:misfit}
\end{center}
\end{figure*}

\begin{figure*}
\begin{center}
\includegraphics[width=\figurewidth]{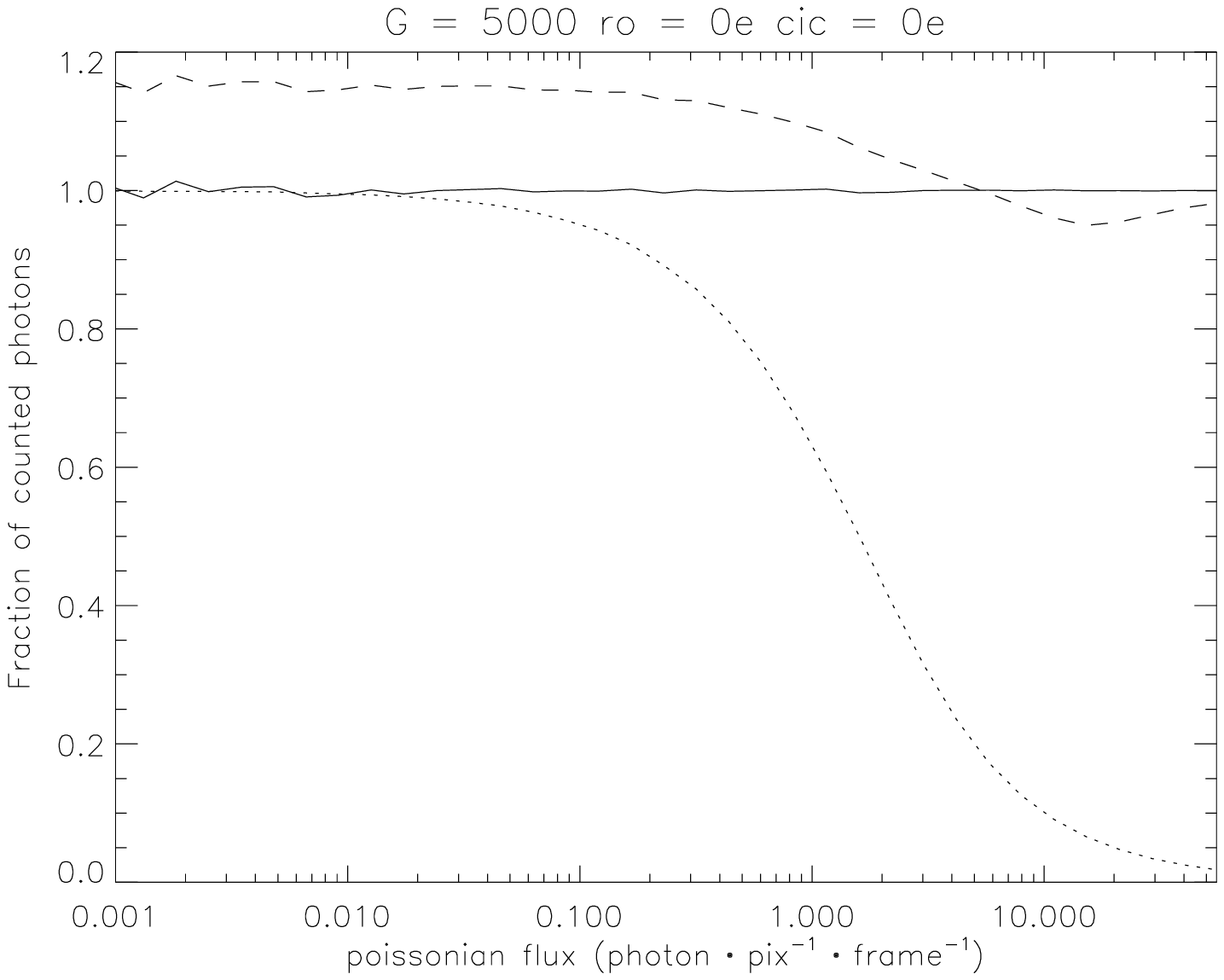}
\includegraphics[width=\figurewidth]{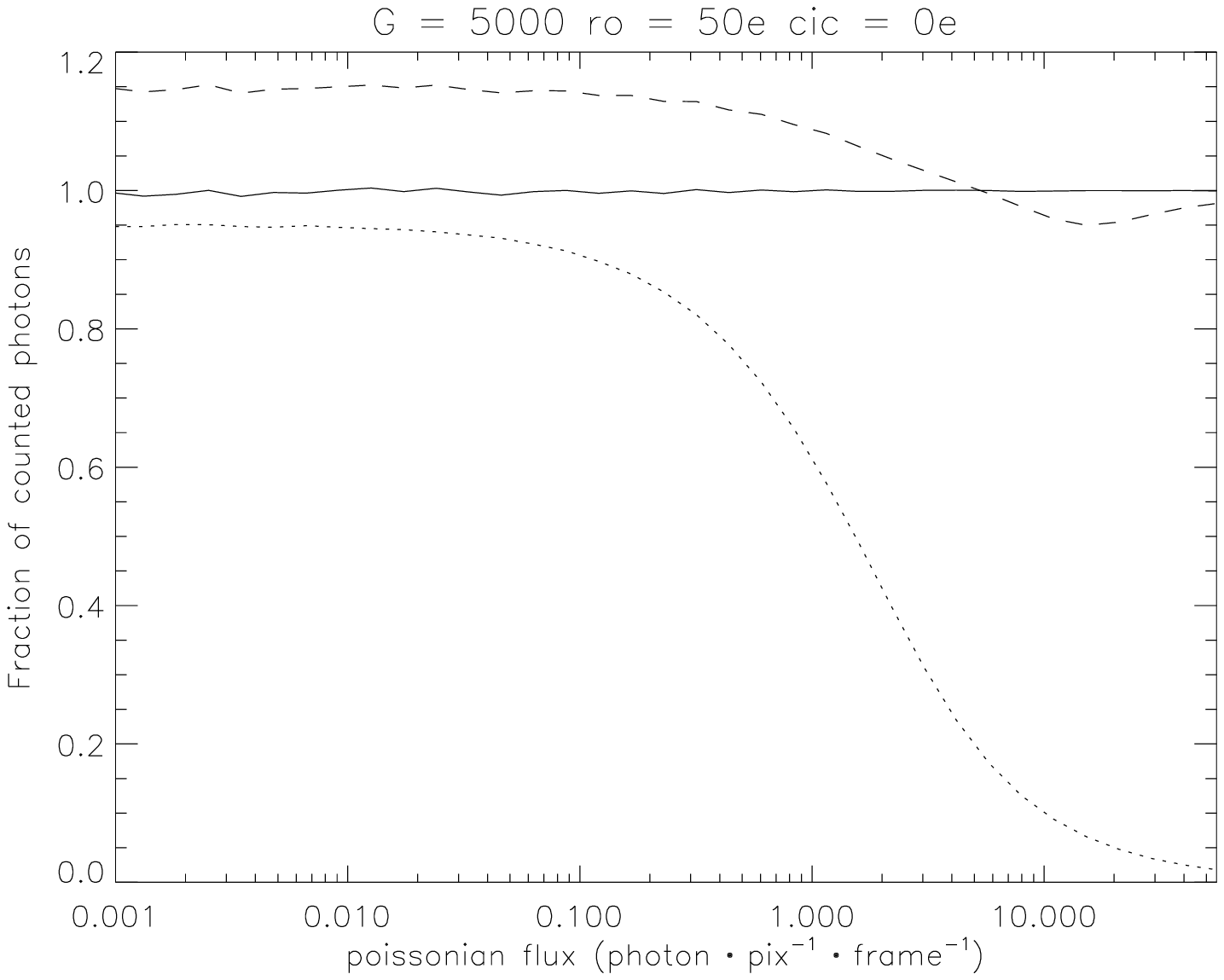}
\caption{Effect of the read-out noise on the fraction of detected photons, for a gain of 5000. Dotted line: PC, dashed line: PP, plain line: analogic.}
\label{fig:fraction}
\end{center}
\end{figure*}

\begin{figure*}
\begin{center}
\includegraphics[width=\figurewidth]{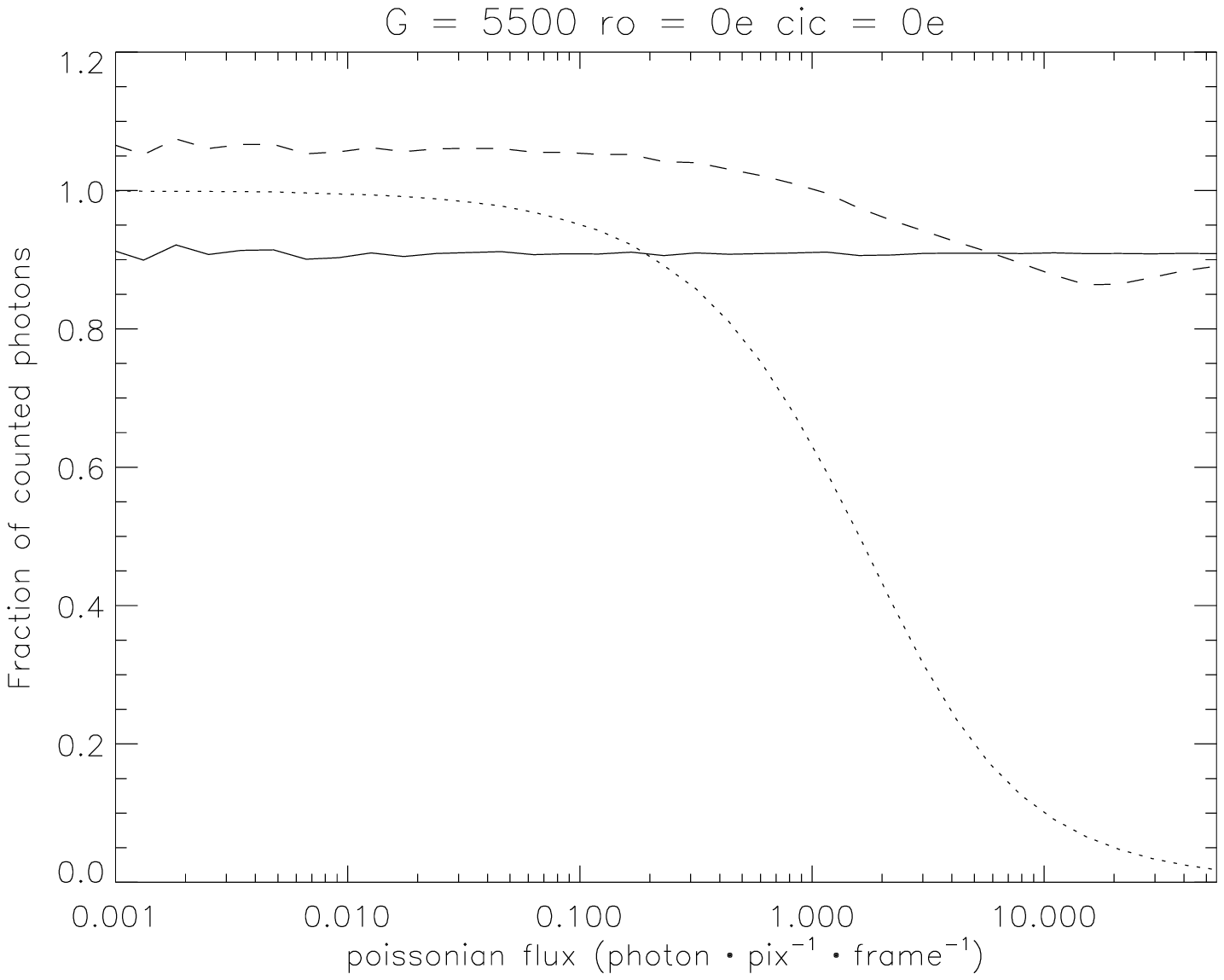}
\includegraphics[width=\figurewidth]{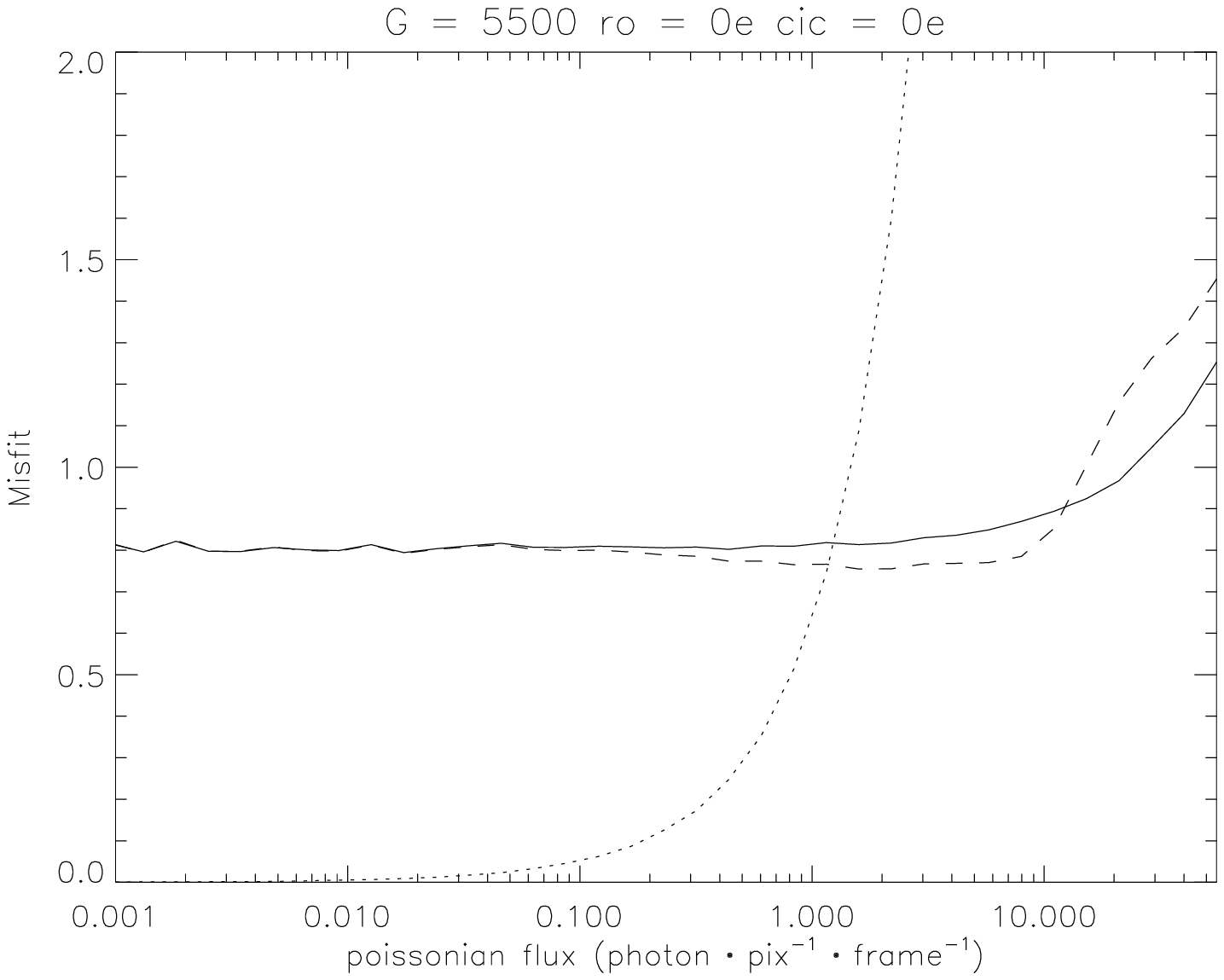}
\caption{Effect of improperly determining the gain (+10\% error) on the fraction of counted photons and on the misfit.}
\label{fig:gain_error}
\end{center}
\end{figure*}

Figure \ref{fig:enf} and the left panel of figures \ref{fig:misfit} and \ref{fig:fraction} show the effect of the different processing methods on the output signal, for a mean gain of 5000. These simulations include neither the read-out noise, nor the generation of spurious charges. The PC processing allows one to get rid of the noise factor induced by the multiplication process at the price of losing events at high fluxes. At low fluxes, this processing induces no error in the output photon count.

The analogic processing behaves as expected: the noise factor of the multiplication process, at high gains, is $\sqrt{2}$.

The PP processing does not produce the expected effect of reducing the noise factor. For fluxes where the photon count per pixel per frame is dominated by single or no event, the ENF is higher than for the pure analogic processing. It is only at a flux between 1 and 10 photons/pixel/frame that there is a slight advantage in using this processing scheme. At any other fluxes, the ENF of the analogic processing is equal or lower.

Regarding the misfit, the curve of the PP thresholding follows the same trend as for the ENF. Thus, the mean counting error of the PP strategy will be lower than the one of the analogic processing for fluxes between 1 and 10 photons/pixel/frame.

Also, figure \ref{fig:gain_error} shows the effect of using a wrong value for the gain. An error of 10\% on the gain value will induce an error of the same magnitude on the fraction of counted photons for the PP and analogic strategies. The misfit will be affected by twice that error. The PC strategy is not affected by the error on the gain as it does not use this value in its algorithm.

\subsection{Effect of the read-out noise}
The effect of the read-out noise, at high gains, is mostly to confuse the thresholding strategies by inducing false events at low fluxes. Thus, a minimum detection threshold must be set to prevent false events from dominating the output. However, since this threshold will induce the loss of the events that would have a low output value, it is important not to set it too high. It was determined in [\citenum{2004SPIE.5499..219D}] that 5.5$\sigma$ was the value that had the less impact on the SNR at faint fluxes, but a lower value could be used at high fluxes to get a very slight gain in SNR. For the simulations, since a read-out noise of 50 electrons is used, the threshold used is set to 275 electrons (5.5$\sigma$). In analogic mode, regardless of the flux or the gain, no thresholding is used as the read-out noise averages to zero. These effects are shown in the right panels of figures \ref{fig:misfit} and \ref{fig:fraction}.

\subsection{Effect of the spurious charges}
\label{sect:cic}

\begin{figure*}
\begin{center}
\includegraphics[width=\figurewidth]{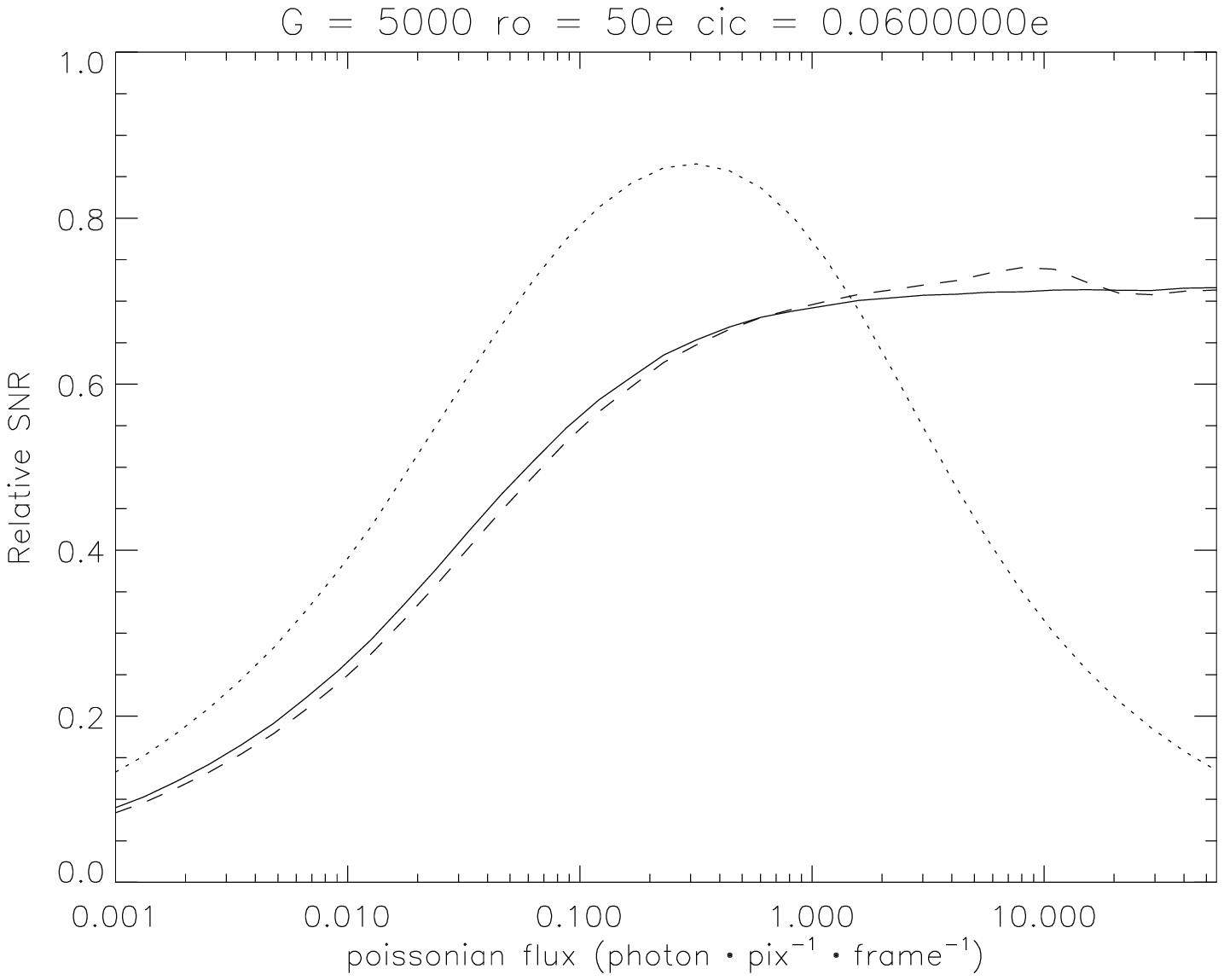}
\includegraphics[width=\figurewidth]{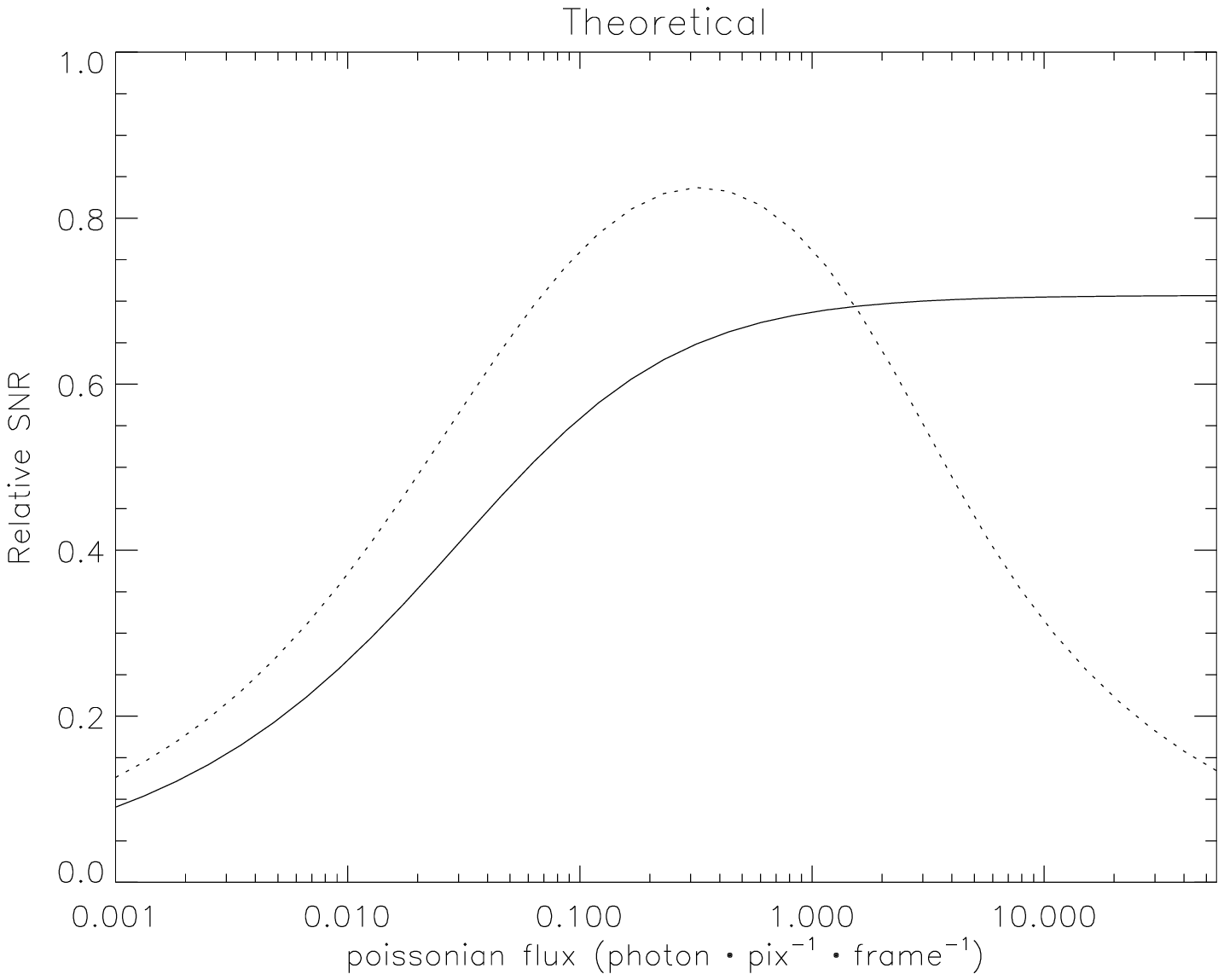}
\caption{Relative SNR of a "real" EMCCD, compared to a perfect photon counting system, including the effect of read-out noise (50 electrons) and the effect of spurious charges (0.06 event/pixel/frame) at a mean gain of 5000. Left panel: result of simulations. Right panel: theoretical values. Dotted line: PC, dashed line: PP, plain line: analogic.}
\label{fig:snr}
\end{center}
\end{figure*}

It is important to add the generation of the spurious charges (or CIC) to properly simulate the behaviour of the SNR of an EMCCD. Most of the spurious charges are generated during the vertical shifting of the charges. The amount of spurious charges generated is also dependant of the operating mode of the CCD (inverted, non-inverted). [\citenum{2004SPIE.5499..219D}] and [\citenum{techreport-minimal}] explain how to choose the best operating mode of the EMCCD regarding the spurious charge generation. However, at low frame rate, it is best to operate the EMCCD in inverted mode, so that the dark noise is diminished.

The simulation results shown in figure \ref{fig:snr} show that by including the CIC and the read-out noise, there is really little difference between the SNR of the analogic processing and the PP thresholding. The highest performance is obtained in photon counting, as this mode does not suffer of the noise factor. Even with a spurious charge rate of 0.06 event/pixel/frame, at a flux of 0.2 photon/pixel/frame the SNR of the EMCCD comes very near the theoretical limit imposed by the shot noise.

The right panel of figure \ref{fig:snr} shows that the value of the SNR of the analogic processing is well explained by this equation
\begin{equation*}
SNR = \dfrac{S}{\sqrt{F^{2}S + F^{2}C + \dfrac{\sigma^2}{G^2}}},
\end{equation*}
where $S$ is the mean flux, $F$ is the noise factor, $C$ is the amplitude of the spurious charge generation, $\sigma$ is the read-out noise and $G$ is the mean gain. For the PC processing, the SNR is expressed as being
\begin{equation*}
SNR = \dfrac{S}{\sqrt{\dfrac{S^2+C^2}{1-e^{S+C}}}}.
\end{equation*}
This last equation reflects the effect on the SNR of counting only one event/pixel/frame in the approximation where the effective read-out noise is $\ll$1 and the threshold causes a negligible loss of events.


\section{Data from an EMCCD}
\label{sect:data}
When it comes to the processing of real data coming out of an EMCCD, a parameter that is taken for granted in the simulations, the mean gain, is not readily available. The mean multiplication gain is in fact dependant of numerous factors, namely the high voltage clock amplitude, the temperature and the age of the EMCCD. The last factor will not vary rapidly. It is only over time that the amplitude of the high voltage phase will have to be slightly increased to maintain the same gain. However, the two other factors will have an impact that may change from frame to frame, and even during a single frame. Even if the amplitude of the high voltage phase is maintained to millivolt accuracy and the temperature is controlled within a fraction of a degree, it is still necessary to develop a way of computing the value of the mean gain from the output of the EMCCD, as it is the key element of an accurate processing of the output (see section \ref{sect:em_gain}). The amplitude of the read-out noise is also a parameter that has to be considered for both PC and PP processing.

\subsection{Determining the amplitude of the read-out noise}
\label{sect:det_ro_noise}
The amplitude of the read-out noise is determined simply by taking many (a hundred or more) dark frames through the EM amplifier set to an EM gain of 1. Then, every pixel is subtracted by the mean value of the corresponding pixels in all the other frames (to overcome the effect of the bias that varies through the frame). The standard deviation of the signal can then be determined to give the amplitude of the read-out noise. The effect of the spurious charges should not affect the computed value. Since a typical value for the read-out noise of the EM amplifier is typically 30-60 electrons (at a high pixel frequency), the noise caused by the spurious charges ($\sim$0.06-0.15 electron/pixel) is completely drown.

This value should not be affected by the temperature of the CCD. Also, since it is only used as the threshold for the detection limit, an error of a few percent will not badly affect the resulting SNR of the system (see section \ref{sect:det_em_gain}).

\subsection{Determining the gain of an EMCCD}
\label{sect:det_em_gain}
The spurious charges that are generated during the vertical transfer can be used to calculate the mean gain of the EMCCD amplification register. Since the intensity of this signal is well below 1 event/pixel/frame, it will be dominated by single events. A thorough understanding of the signal processing can be used to recover the value of the gain. The iterative process described below can be used.

As a first approximation, the gain is estimated by finding the mean value of the pixels that have a value greater than the read-out noise. The value of the read-out noise determined in section \ref{sect:det_ro_noise} can be used with a 5.5$\sigma$ threshold. Then, one can estimate the signal level (the amount of spurious charges per pixel per frame), $f$, that is seen on the CCD by simply dividing the mean value of all the pixels by the gain found. These two approximated values for the gain and the signal, are then refined by mean of an iterative process:

\begin{enumerate}
\item The threshold applied will cause the loss of the events that have an output value lower than the threshold. One can estimate the amplitude of this loss by finding the probability of having output values lower than the threshold for the estimated gain, assuming that pixels will not have undergone more than one event:
\begin{equation*}
p(x<th) = \sum_{x=1}^{th}{\dfrac{e^{-x/g}}{g}},
\end{equation*}
where $th$ is the threshold expressed in electrons and $g$ is the approximated gain. The estimated signal level may then be corrected for these lost events with
\begin{equation*}
f_{1} = \dfrac{f}{1-p(x<th)},
\end{equation*}
where $f$ is the approximated signal level.

\item Then, one may also remove the events that were counted as real events but were due to the read-out noise. This is simply the probability of having events of an amplitude higher than $5.5\sigma$ for a normal distribution of standard deviation $\sigma$, which is
\begin{equation*}
p(x>5.5\sigma) = 0.5\left[1-\mathrm{erf}\left(\dfrac{5.5}{\sqrt{2}}\right)\right],
\end{equation*}
where $\mathrm{erf}$ is sometimes called the \textit{error function} and is defined as being
\begin{equation*}
\mathrm{erf}(z) = \dfrac{2}{\sqrt{\pi}}\int_0^z{e^{-t^2}dt}.
\end{equation*}
Then, one can once more correct the signal amplitude with
\begin{equation*}
f_{2} = f_{1} \left(1-p(x>5.5\sigma)\right),
\end{equation*}
where $f_{1}$ is the signal amplitude found at step 1. This correction is very small for a threshold value of $5.5\sigma$. However, if a lower threshold is used, it will become increasingly important to apply it.

\item So far, calculations involved assuming that no more than one event per pixel was generated. However, even at low signal values, there are slight chances that a Poissonian process will generate more than one event into a single pixel. This can be compensated with the following equation
\begin{equation*}
g = \dfrac{1-e^{-\alpha}}{\alpha},
\end{equation*}
which gives $g$, the proportion of counted events for a Poissonian process generating $\alpha$ events per interval. Since the exact value of the signal level is not known, one may simply use $\alpha = f_2$ when $f_2$ is small ($<0.4$) to approximate $g$. Then, the signal amplitude, may be corrected with
\begin{equation*}
f_{3} = \dfrac{f_{2}}{g},
\end{equation*}
where $f_{2}$ is the value calculated at step 2.

\item Finally, a new gain may be calculated by using the corrected signal amplitude of step 3 and by simply dividing the mean value of all the pixels by this new signal amplitude. Then, one may loop to step 1 to calculate an increasingly accurate gain, using the gain just calculated as the input of the algorithm. However, the signal amplitude estimate $f$, used to iterate, must be the one calculated as a first approximation (rather than $f_3$). 
\end{enumerate}

This algorithm can determine the gain of the EMCCD to a fraction of a percent accuracy. In fact, the achievable accuracy is of the order of
\begin{equation*}
\eta \simeq \sqrt{\dfrac{\sigma}{n_c}},
\end{equation*}
where $n_c$ is the total detected charges and $\sigma$ is the read-out noise. Thus, if the spurious charge generation rate is of the order of 0.1 event/pixel/frame, the read-out noise is of the order of 50 electrons and if one wants to determine the gain to an accuracy of 0.1\%, a \textit{strict minimum} of $5 \times 10^8$ pixels must be used, which means the processing of $\sim$2000 full 512x512 dark frames.

The exact knowledge of the amplitude of read-out noise is not necessary to determine the gain. Through simulations, it was determined that an error of 10\% on the amplitude of the read-out noise induces an additional relative error of the same amplitude. Thus, for a wanted accuracy of 0.1\%, an error of 10\% on the read-out noise will give a final accuracy of 0.11\%. However, a greater error on the determination of the read-out noise will induce an ever increasing error on the gain as the algorithm will get confused. Also, this algorithm is working fine for systems where the gain is more than 10 times larger than the read-out noise.

This algorithm assumes that the spurious charges are generated before entering the EM register. Thus, if some of the charges are generated into the EM register, all charges will not travel the same amount of multiplication elements and will therefore not have the same mean gain at the output. However, the algorithm described can still be used to test the variability of the gain of an EMCCD as the amplitude of the noise arising from the serial register should be steady (in a temperature controlled environment) and will cause the gain to be under-estimated by a fixed ratio. Thus, this code can be used to determine \textit{in-situ} if the gain of the EM stage has changed. If the amount of spurious charges generated into the serial register is known, the exact gain can still be extracted from this algorithm by applying a conversion factor.

\section{Conclusions}
The numerical simulations of the EMCCD show that the noise factor induced by the statistical gain cannot be overcome by means of the PP processing explained in [\citenum{2003MNRAS.345..985B}]. The simulations also show that the EMCCD is behaving as expected in analogic mode, where the noise factor of $2^{1/2}$ has the effect of halving the QE of the CCD. The PC processing allows one to overcome the noise factor at the price of detecting no more than one event/pixel/frame.

The accurate determination of the gain of the EMCCD is possible through the algorithm developed in \ref{sect:det_em_gain}. An IDL routine implementing this algorithm is available at \url{http://www.astro.umontreal.ca/~odaigle/emccd}.

\bibliography{spie_l3_2006}
\bibliographystyle{spiebib}
\clearpage
\end{document}